\documentclass{article}

\usepackage{epsfig}
\usepackage{amsmath}
\usepackage{amssymb}
\usepackage{bm}
\usepackage{multirow}
\usepackage{booktabs}
\usepackage{epstopdf}
\usepackage{subcaption}
\usepackage{graphicx}
\usepackage{authblk}

\newcommand{\mtr}[1]{\bm{#1}} 
\newcommand{\tran}[1]{{#1}^{\text{T}}}   

\makeatletter

\begin{document}

\title{A Method for Neuronal Source Identification}
\author[1]{Chang Won Lee}
\author[2]{Agnieszka A. Szymanska}
\author[3]{Shun Chi Wu}
\author[4]{A. Lee Swindlehurst}
\author[2,4]{Zoran Nenadic} 
\affil[1]{Samsung Dallas Technology R\&D Lab, Richardson, TX, USA}
\affil[2]{Department of Biomedical Engineering, University of California, Irvine, CA, USA}
\affil[3]{Covidien, Costa Mesa, CA, USA}
\affil[4]{Department of Electrical Engineering and Computer Science, University of California, Irvine, CA, USA}
\date{}
\maketitle

\begin{abstract}
Multi-sensor microelectrodes for extracellular action potential recording have significantly improved the quality of {\it in vivo} recorded neuronal signals. These microelectrodes have also been instrumental in the localization of neuronal signal sources. However, existing neuron localization methods have been mostly utilized {\it in vivo}, where the true neuron location remains unknown. Therefore, these methods could not be experimentally validated. This article presents experimental validation of a method capable of estimating both the location and intensity of an electrical signal source. A four-sensor microelectrode (tetrode) immersed in a saline solution was used to record stimulus patterns at multiple intensity levels generated by a stimulating electrode. The location of the tetrode was varied with respect to the stimulator. The location and intensity of the stimulator were estimated using the Multiple Signal Classification (MUSIC) algorithm, and the results were quantified by comparison to the true values. The localization results, with an accuracy and precision of $\sim$10 $\mu$m, and $\sim$11 $\mu$m respectively, imply that MUSIC can resolve individual neuronal sources. Similarly, source intensity estimations indicate that this approach can track changes in signal amplitude over time. Together, these results suggest that MUSIC can be used to characterize neuronal signal sources {\it in vivo}.

\end{abstract}



\section{Introduction} \label{sec:intro}
Multi-sensor microelectrodes, such as tetrodes, are increasingly being used for the extracellular recording of action potentials (APs). In addition to improving AP detection and classification by increasing the signal-to-noise ratio (SNR)~\cite{Gray1995}, multi-sensor probes can be used to localize and characterize single neurons~\cite{Jog2002, Aur2005, Mechler2011}. This function has, however, been largely unexamined and remains underutilized in experimental neuroscience. As more studies are performed using these probes, the need to localize and characterize neurons based on recorded APs is becoming increasingly important in deducing neuronal function. 

The prospect of localizing neural signal sources has specific benefits for both acute and chronic extracellular recoding experiments. Electrode positioning and guidance during acute recording is a very involved, time consuming process. Furthermore, electrode micromanipulators have only one degree of freedom, and data is often lost as neurons migrate away from the probe track. If these migration trends could be estimated over a short period of time, experimentalists could determine if a neuron is a good candidate for further analysis or likely to travel away from the probe tack. Additionally, autonomous algorithms for single neuron isolation and tracking~\cite{Nenadic2006,Cham2005,Burdick2012} could be greatly improved with neuron location data. In chronic recording, localization can be used to track the migration trends, as well as firing patterns of neurons and neuronal populations. This could have significant implications for the study of neural plasticity~\cite{Kuboshima-Amemori2007}, cell assembly functional organization~\cite{Harris2003}, scar tissue formation~\cite{Polikov2005}, as well as neural network connectivity and communication~\cite{Bartho2004}. 

The intensity of the extracellular potential field created when a neuron fires an AP is another important neuronal characteristic that is often overlooked due to inadequate data processing techniques. 
As large neurons emit a stronger signal, they are often over-represented in population studies~\cite{Humphrey}. Currently it is not possible to distinguish a small nearby neuron from a large distant one. The ability to determine a neuron's transmembrane current intensity as well as location may lead to better understanding of size-related neural functional properties.

This article presents a statistical signal processing technique suitable for localization and current intensity estimation of single neuronal sources, based on multi-sensor measurements of their extracellular APs. To prove the validity of our technique beyond {\it in silico} models, a simple experiment was performed testing the algorithm's ability to localize and estimate the strength of an artificially generated monopole-like source. Stimulating signals were generated at various strengths and recorded with a commercial tetrode. The success of our technique makes it a good candidate for further {\it in vitro} testing and {\it in vivo} use. 

\section{Preliminary Work} \label{sec:in_silico} 
Source localization is an important problem in experimental neuroscience, and several methods have been developed to localize source neurons during extracellular recording~\cite{Mechler2011, Bartho2004, Csicsvari2003, Chelaru2005, Somogyvari2005}. However, most of these methods~\cite{Bartho2004, Csicsvari2003, Somogyvari2005} are not capable of localizing sources in three-dimensional (3D) space. Furthermore, as these were {\it in vivo} studies, none of these methods were fully validated. Our preliminary work has shown that a closed-form solution to the neuronal source localization problem can be found given a 3D sensor arrangement, and assuming a monopole forward model~\cite{Lee2007}. This method, while relatively tractable, is sensitive to noise, and therefore, a solution may not always exist. Additionally, the geometry of the problem always leads to two solutions. Most often, the spurious solution lies inside the sensor array and can be easily identified and discarded. However, sometimes the two solutions cannot be disambiguated. Due to these limitations another method relying on statistical signal processing was considered and tested {\it in silico} against the closed-form solution. This statistical signal processing method relies on the Multiple Signal Classification (MUSIC) algorithm~\cite{rschmidt:86}, which has been previously used in electroencephalogram~\cite{Wu2012, Wu2012a} and magnetoencephalogram~\cite{Mosher1998} source localization. MUSIC therefore presents a promising technique for neuronal source localization. 

\begin{figure}[!htbp]
	\centering
	\begin{subfigure}[t]{0.45\textwidth}
		\centering
		\includegraphics[width=2.9in]{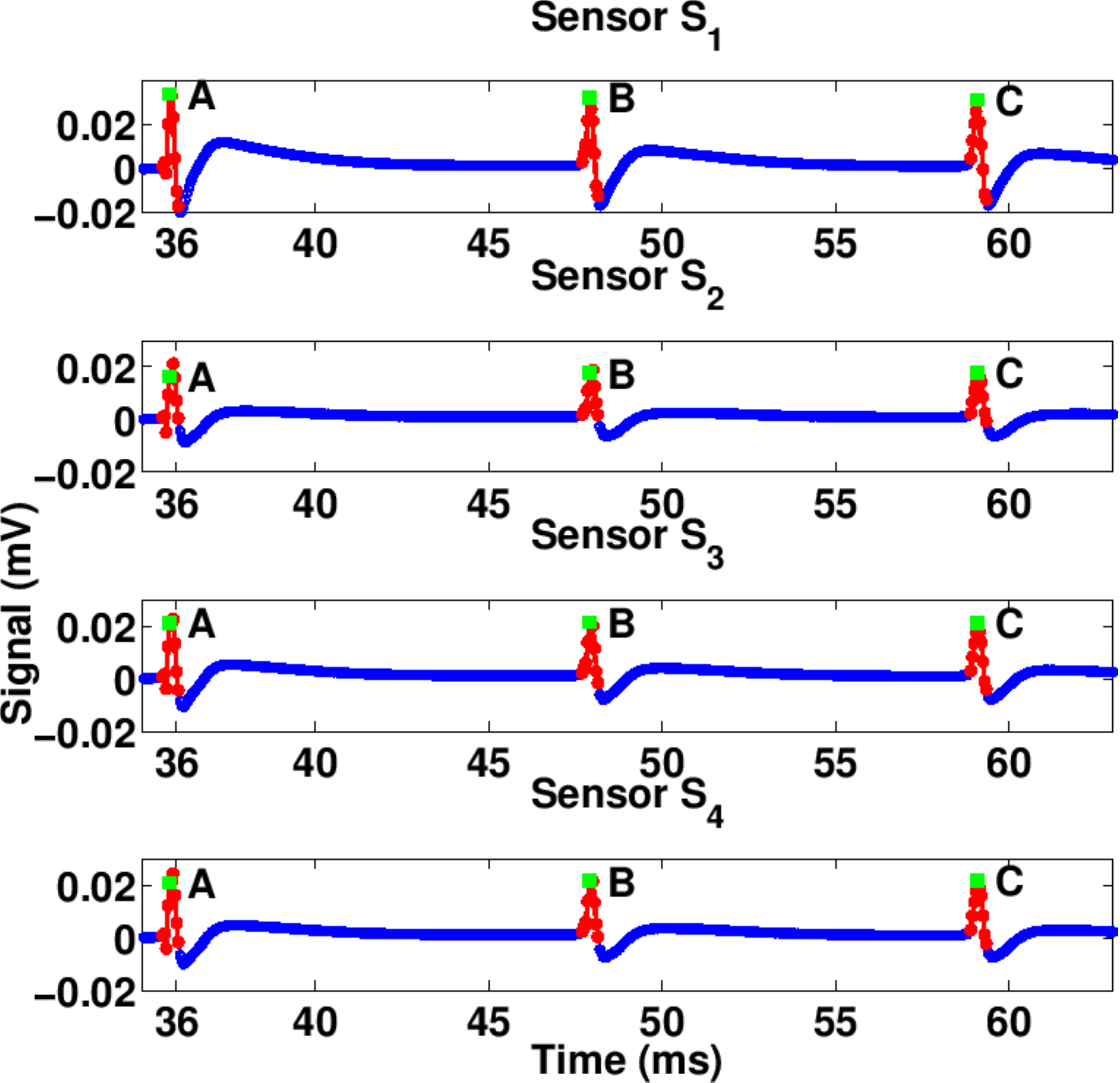}
		\caption{}
		\label{fig:silico_sig}
	\end{subfigure}%
	\hspace{0.01in}
	\begin{subfigure}[t]{0.54\textwidth}
		\centering
		\includegraphics[width=3.5in]{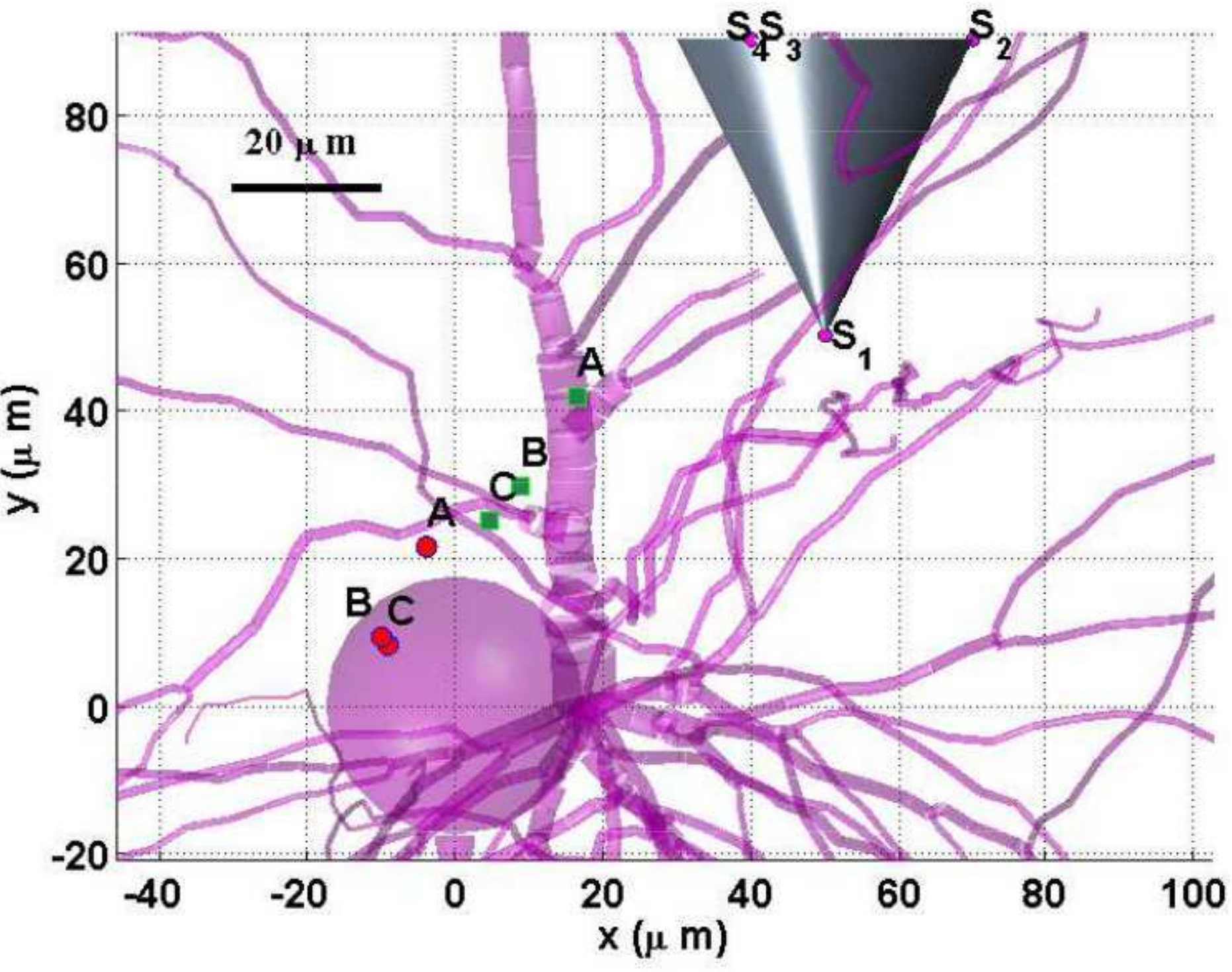}
		\caption{}
		\label{fig:silico_res}
	\end{subfigure}
	\caption{Comparison of closed-form and MUSIC localization solutions. (a) Three differently shaped APs, labelled A, B and C, were generated for all four sensors of a simulated tetrode, and used to localize the source. When calculating source locations, the closed-form approach uses only one data point from the AP peak, marked in green, while MUSIC takes as input a time series describing the AP, marked in red. (b) Localization results for each AP, labelled A, B, and C, plotted with the model neuron. The simulated tetrode and neuron soma are potted respectively as a silver cone with sensors $S_{1-4}$, and a purple sphere. Closed-form and MUSIC derived localization solutions are shown as green and red points respectively.}
	\label{fig:silico}

\end{figure}

To underscore the differences between the two methods, we compared closed-form and MUSIC-derived solutions, based on a computational 
model of a neuron~\cite{Mainen1996}. Using this model, three differently shaped APs of a bursting neuron, labelled A, B and C in Fig.~\ref{fig:silico_sig}, were generated at four locations 
 arranged to mimic a commercially available tetrode (Thomas Recording, Geissen, Germany). These signals were used for localization. The closed-form as well as MUSIC-derived localization results are shown in Fig.~\ref{fig:silico_res}. The closed-form approach uses a single amplitude measurement, marked in green on the three APs in Fig.~\ref{fig:silico_sig}, to estimate source locations, plotted as green squares in Fig.~\ref{fig:silico_res}. Contrarily, the MUSIC algorithm takes as input a time series describing the shape of the AP, as outlined in red in Fig.~\ref{fig:silico_sig}. The corresponding calculated source locations are plotted as red circles in Fig.~\ref{fig:silico_res}. The MUSIC algorithm, as seen in Fig.~\ref{fig:silico}, better estimated the location of the simulated neuron, and was therefore chosen for further validation {\it in vitro}. 

\section{Materials and Methods}\label{sec:mam}


As an action potential propagates, positive sodium ions enter the neuron through voltage gated ion channels. This process starts near the soma, and incrementally travels down the axon and up the dendritic tree, causing the neuron to polarize. The neuron then re-establishes equilibrium by actively pumping out the sodium ions. In this way, the neural membrane can be described as a distributed, dynamic network of ion current sources and sinks. However, this multiple source and sink model of the neural membrane proves to be overly complex for localization purposes, prompting the need for a simpler approximation. In the most basic case, we can treat the neuron as a point source and the surrounding medium as an isotropic, homogeneous volume conductor. Although simplistic, the monopole model has been used in application to neural source localization~\cite{Csicsvari2003, Bartho2004, Aur2005, Chelaru2005}. The potential at a specific sensor $S_{i}$ is then 
\begin{equation}
\psi_{i} = \frac{I(t)}{4 \pi \sigma d_{i}(\mtr{r})} \label{eq:monopolemodel}
\end{equation}
where $I(t)$ is the time varying source current, $\sigma$ is the conductivity of the medium per unit length, and $d_{i}$ is the distance between the sensor $S_{i}$ and the source located at $\mtr{r}=\tran{[x,\,y,\,z]}$. We will use this as the basis of our forward model. However, given that a homogeneous volume conductor is difficult to generate experimentally, the behavior of the model when this assumption is violated will also be addressed. 

\subsection{Multiple Signal Classification Algorithm}\label{sec:msca}
If signals are generated by a single source, the MUSIC algorithm models measurements from a $c$-sensor array, $\mtr{\psi}(t)\in\mathbb{R}^{c\times 1}$, as an output of the static linear system
\begin{equation}\label{eq:model}
\mtr{\psi}(t)=\mtr{m}{s}(t)+\mtr{w}(t)
\end{equation}
where $t$ is the time instant, $\mtr{m}\in\mathbb{R}^{c\times 1}$ is the lead field vector (LFV)~\cite{Wu2012, Wu2012a} representing the system's response to a unitary signal input, ${s}(t)\in \mathbb{R}$ is the signal  amplitude, and $\mtr{w}(t)\in\mathbb{R}^{c\times 1}$ is zero-mean noise. In the case of a single monopole-like source with current $I(t)$, the LFV becomes
\begin{equation}\label{eq:lfv}
\mtr{m}(\mtr{r}) = \frac{1}{4\pi \sigma}\tran{\left[ \frac{1}{d_1(\mtr{r})} \, \frac{1}{d_2(\mtr{r})} \, \cdots \, \frac{1}{d_c(\mtr{r})} \right]}
\end{equation}
and $s(t)=I(t)$. Eq.~(\ref{eq:model}) then takes the form of the initial forward model with added zero mean noise. The components of $\mtr{\psi}(t)$ for any sensor $i \in [1, 2, \cdots, c]$ are then 
\begin{equation} 
\psi_{i} = \frac{I(t)}{4 \pi \sigma d_{i}(\mtr{r})} + w_i(t) \label{eq:monopolwnoise}
\end{equation}
The MUSIC algorithm proceeds by finding the source location $\mtr{r}^{\star}$ for which the LFV is most orthogonal to the noise subspace~\cite{rschmidt:86}. More formally, the optimal source location $\mtr{r}^{\star}$ is found by
\begin{equation}\label{eq:music}
\mtr{r}^{\star} = \arg\min_{\mtr{r}}\frac{\tran{\mtr{m}}(\mtr{r})\mtr{E}_N\tran{\mtr{E}}_N\mtr{m}(\mtr{r})}{\tran{\mtr{m}}(\mtr{r})\mtr{m}(\mtr{r})}
\end{equation} 
where $\mtr{E}_N\in\mathbb{R}^{c\times (c-1)}$ is the noise subspace. This subspace can be obtained by the following singular value decomposition,
\begin{equation}\label{eq:svd}
\mtr{\Psi} = \mtr{U}\mtr{S}\tran{\mtr{V}}
\end{equation}
where $\mtr{\Psi}:=\left[\mtr{\psi}(1) \,\mtr{\psi}(2) \, \cdots \, \mtr{\psi}(T)\right]\in\mathbb{R}^{c\times T}$ and $T$ is the number of samples in the time series data. If $T\ge c$ under the single-source assumption, the noise subspace can be defined as $\mtr{E}_N:=\left[\mtr{u}(2) \,\mtr{u}(3) \,\cdots \,\mtr{u}(c)\right]$, where $\mtr{u}$ represents the columns of $\mtr{U}$ corresponding to the $c-1$ smallest singular values of $\mtr{\Psi}$. In other words, we assume that the first singular value of $\mtr{\Psi}$  makes up the signal subspace, and the remaining values make up the noise subspace $\mtr{E_N}$. Note that $\sigma$ cancels out in~(\ref{eq:music}) so localization is independent of medium conductivity. 

\subsection{Inhomogeneity Correction Factor}\label{sec:icf}
In general, the conducting medium may not be homogeneous and isotropic, which implies that $\sigma$ in Eq.~(\ref{eq:lfv}) may not be constant. Allowing each source-sensor path to have its own conductivity value redefines the LFV as
\begin{equation}\label{eq:lfvm}
{\mtr{\bar{m}}}(\mtr{r}) = \frac{1}{4 \pi \sigma}\tran{\left[ \frac{1}{k_1 d_1(\mtr{r})} \, \frac{1}{k_2 d_2(\mtr{r})} \, \cdots \, \frac{1}{k_c d_c(\mtr{r})}\right]}
\end{equation}
where $k_i$, $i \in [1, 2, \cdots, c]$, is a constant making each conductivity a multiple of some baseline value $\sigma$. To determine $k_i$, referred to as the inhomogeneity correction factor (ICF), we set the conductivity of the sensor closest to the source, as this baseline, $\sigma_{i^{\star}}=\sigma$ and therefore $k_{i^{\star}}=1$. After combining Eqs.~(\ref{eq:model}) and~(\ref{eq:lfvm}), and taking the expectation over the zero mean noise distribution, we obtain 

\begin{equation}\label{eq:expectation_val}
E\{\psi_i(t)\}=\frac{I(t)}{4\pi k_i\sigma d_i}, \quad i = 1,\,2,\,\cdots,\,c 
\end{equation}
which can be used to solve for $k_i$~\cite{Lee2011} 
\begin{equation}\label{eq:icf}
k_i = \frac{d_{i^{\star}}E\{\psi_{i^{\star}}(t)\}}{d_{i}E\{\psi_{i}(t)}, \quad i = 1,\,2,\,\cdots,\,c \quad (i\ne i^{\star})
\end{equation}
The forward model~(\ref{eq:model}) can then be rewritten as
\begin{equation}\label{eq:modelm}
\bar{\mtr{\psi}}(t)=\mtr{m}(\mtr{r})\mtr{s}(t)+\bar{\mtr{w}}(t)
\end{equation} 
where 
the components of $\bar{\mtr{\psi}}$ and $\bar{\mtr{w}}$ are  $\bar{\psi}_i(t) = k_i\psi(t)$, and $\bar{w}_i(t) = k_i w(t)$ respectively. The MUSIC algorithm [Eqs.~(\ref{eq:model}-\ref{eq:svd})] can then be executed with $\bar{\mtr{\psi}}$ instead of $\mtr{\psi}$.    

\subsection{Accuracy and Precision}\label{sec:aap}
The accuracy of an estimate is measured by the difference between the estimated value of the parameter and its true value 
. In the case of 3D source localization, this difference, also referred to as an error or bias, is defined as
\begin{equation}\label{eq:error}
\varepsilon := \|\mtr{r}_e-\mtr{r}_t\|
\end{equation}  
where $\|.\|$ represents the Euclidean norm, and $\mtr{r}_e, \mtr{r}_t\in \mathbb{R}^{3\times 1}$ are the locations of the estimated and true source, respectively. 

The precision of an estimate is measured by its spread (standard deviation) around its mean value 
. To generalize this notion to multivariate parameters, we define a standard radius as
\begin{equation}\label{eq:sr}
\delta := \sqrt{\delta_x^2+\delta_y^2+\delta_z^2}
\end{equation}   
where $\delta_x, \delta_y$ and $\delta_z$ are the standard deviations of the $x, y$ and $z$ components of $\mtr{r}_e$, respectively. 

\subsection{Experimental Setup}\label{sec:es}

\begin{figure}[!htbp]
	\centering
		\includegraphics[width=0.9\textwidth]{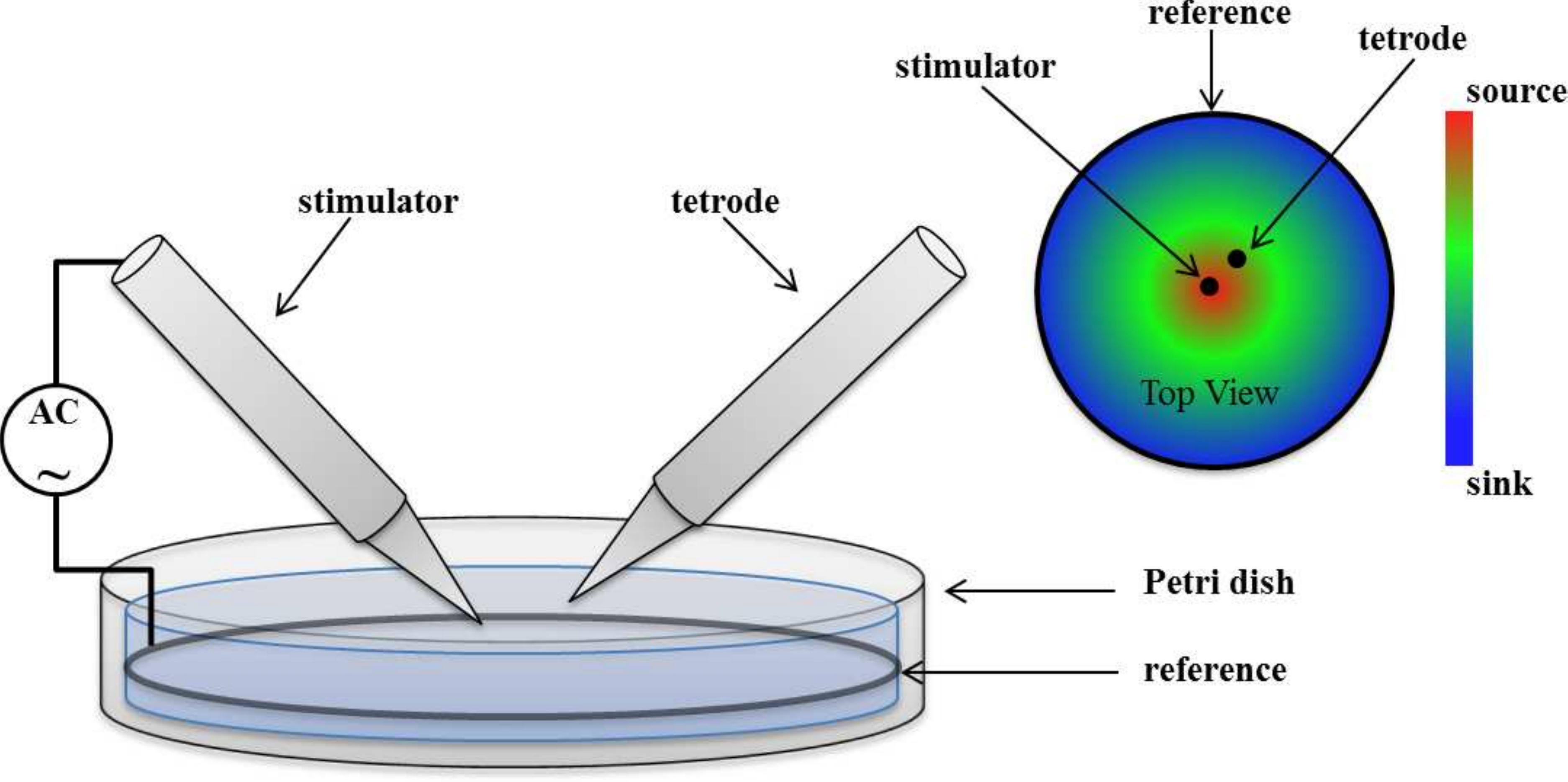}
	\caption{Schematic of experimental setup. (Left) A stimulator (metal microelectrode) connected to a voltage source with a circular reference. The stimulator is placed in a Petri dish with 0.9\% saline solution.  A tetrode is placed in close proximity to the stimulator. (Right) A top view of the setup with the assumed electric potential field.}
	\label{fig:schematic}
\end{figure}

To test the performance of our method on experimental data, a metal microelectrode (Alpha Omega Co. USA, Alpharetta, GA) was connected to an AC voltage source (MP150, Biopac, Goleta, CA), with a circular reference electrode made out of 0.20 mm$^2$ wire. The microelectrode, which served as a stimulator, was placed in the center of a Petri dish filled with 0.9\% (by volume) saline solution (Fig.~\ref{fig:schematic}). The reference electrode was placed against the Petri dish wall along its circumference. Due to radial symmetry, this electrode arrangement produces a monopole-like electric potential field. The voltage source was then programmed to produce a 7-Hz sine wave at amplitudes of 0.5, 0.7, 1.0, 1.2, and 1.5 V. This stimulation frequency was chosen because it is not a subharmonic of power line noise (60 Hz), thus facilitating unambiguous signal detection. The amplitude of the stimulus was varied to additionally  test the algorithm's ability to estimate the intensity of the source signal. Field potentials were then measured by bringing a tetrode (Thomas Recording, Geissen, Germany) in close proximity to the stimulator. The tetrode was positioned under microscope guidance (Olympus IX51, Olympus America, Center Valley, PA) using a combination of course and fine movements performed by a micromanipulator  (Narishige International USA, East Meadow, NY) and a motorized microdrive (MiniMatrix, Thomas Recording, Geissen, Germany) respectively. The position was deemed satisfactory once the tetrode tip was in the same microscope focal plane as the stimulator tip. 

For each stimulation amplitude, 100 cycles of the sine wave were generated, and then collected, by the tetrode. The setup was connected to an optically isolated data acquisition system (RX7, Tucker-Davis Technologies, Alachua, FL). Signals were sampled at 25 KHz, digitized at 16-bit resolution, low-pass filtered with a cut-off frequency of 3000 Hz, and notch filtered to eliminate 60 Hz noise. After recording, the collected 100 cycles were aligned and averaged for each stimulating amplitude. All experiments were performed at room temperature (20$^{\circ}$C). 

\subsubsection{Virtual Sensor Arrays}\label{sec:methodVT}
Measurements were taken at five different tetrode positions, as shown in Fig.~\ref{fig:vt}. The tetrode position was changed using a coarse movement micromanipulator, while keeping the tetrode tip and the stimulator in the same focal plane. This approach was necessary in order to precisely estimate the position of the sensors with respect to the stimulator. Note that based on microscope images, only the location of the  tip sensor can be determined unambiguously, while the other three sensors are in general not visible. The MUSIC algorithm can be used to determine the relative locations of the remaining three sensors, based on knowledge of the tip sensor location and the manufacturer's specifications (Sec.~\ref{sec:d_sensor_locations}). However, this would confound the validation process. To make sure all sensor locations are precisely identified, a virtual sensor array was constructed. The five tetrode-position setup allows for multiple sensor arrangements to be considered. Combining measurements from four out of five tip sensors defines a tetrode-like arrangement and will be referred to as a virtual tetrode (VT). Specifically, the arrays consisting of tip sensors 1, 2, 4, and 5, and 2, 3, 4, and 5, will be referred to as VT1 and VT2, respectively (Fig.~\ref{fig:vt}). Similarly, to ascertain whether performance is improved by the addition of sensors, data from all five tip sensors was combined into a virtual pentode (VP) array. The distances between the stimulator tip and the tetrode tips at positions 1 - 5 were 163.6,	94.6,	95.5,	38.3, and	81.2 $\mu$m, respectively. This experimental setup ensures that all ``ground truth'' parameters are precisely estimated, including the relative locations of the stimulator and the sensors in each virtual array, allowing the MUSIC algorithm's performance to be properly assessed. Note that since only one tetrode was used for recording, multi-sensor VT and VP data could not be acquired simultaneously.  

%

\section{Results}\label{sec:r}
The MUSIC algorithm was used to estimate source locations given collected signals. The localization solutions were then used to estimate source current amplitudes. MUSIC source localization results showed the algorithm's ability to accurately and precisely predict the true source location. However, analysis also indicated that the common assumption of a homogeneous medium did not hold, even in the simple saline medium used here. Current amplitude calculations proved that our method was also capable of accurately tracking changes in source intensity. These results imply our algorithm's ability to resolve small nearby neurons form large distant ones, which may help uncover size-specific neuronal functions. 

\subsection{Source Localization}

\subsubsection{Virtual Tetrode} \label{sec:vt}
\begin{figure}[!htbp]
	\centering
		\includegraphics[width=0.5\textwidth]{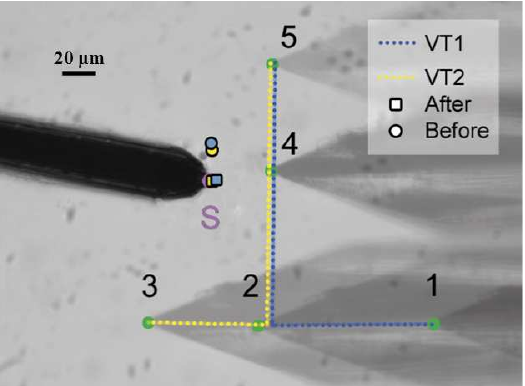}
	\caption{Superposition of five microscope images showing five tetrode positions (1 - 5) with tip positions marked by green circles. The stimulating electrode is on the left with the source, S, defined at the tip and marked by a pink circle. The tetrode and stimulator were placed in the same focal plane for each depicted position. MUSIC estimated source locations, averaged over 5 stimulus amplitudes, are shown as blue (VT1) and yellow (VT2) circles. Blue and yellow squares show the same parameters estimated after accounting for medium inhomogeneity.}\label{fig:vt}
\end{figure}

The performance of the algorithm was first tested based on data from two virtual tetrodes, VT1 and VT2. These two configurations shared data from tip sensors 2, 4, and 5. However, due to its use of sensor 3 as opposed to sensor 1, VT2 was relatively closer to the stimulator than VT1 (Fig.~\ref{fig:vt}).  

Aligned and averaged data was used as input to the MUSIC algorithm for each stimulation amplitude, yielding  estimated source locations. The stimulator tip S was defined as the origin. Table~\ref{tab:vt_before} shows these parameters, as well as the average location of the estimated sources for both VT1 and VT2. For VT1, the estimated source error ranged between $\sim$19 $\mu$m and $\sim$25 $\mu$m, with the average estimated source location (blue circle in Fig.~\ref{fig:vt}) being 21.52 $\mu$m from the source. The error did not appear to correlate with the stimulation amplitude ($\rho=-0.16$, $\text{p-value} = 0.81$). For VT2, the estimated source error ranged between $\sim$15 $\mu$m and $\sim$44 $\mu$m, with the average estimated source location (yellow circle in Fig.~\ref{fig:vt}) being 18.40 $\mu$m from the source. In this case, the error was inversely correlated with the stimulation amplitude ($\rho=-0.97$, $\text{p-value} = 0.0063$). Another difference between the two architectures was a lower precision for VT2 ($\delta=24.86$ $\mu$m) than VT1 ($\delta=4.10$ $\mu$m).
\begin{table}[!htbp]%
\caption{Estimated source locations based on data acquired for VT1 and VT2 at different stimulation amplitudes. Estimated source location error, $\varepsilon$, is defined in Eq.~(\ref{eq:error}), with the true source located at the origin, $\mtr{r}_t:=\tran{\left[0,\,0,\,0\right]}$. The last two columns show the average estimated source location and its standard deviation, respectively, across all stimulation amplitudes. The standard radius, $\delta$, was calculated using Eq.~(\ref{eq:sr}).}
\label{tab:vt_before}
\begin{center}
\begin{tabular}{crrrrrrrr}
\multicolumn{2}{c}{Stimulus amp. (V)} & 0.5 & 0.7 & 1.0 & 1.2 & 1.5 & Ave. & Std. dev.\\ \toprule
\multirow{4}{*}{VT1\hspace{0.4in}} & $x$ ($\mu$m) &-20.04&-22.10 &-25.11&-22.45&-17.68&-21.48&2.78\\
 		& $y$ ($\mu$m) & 0.85	& 0.82 & 0.10 &	-1.03 &	-0.56	& 0.04 & 0.83 \\
 		& $z$ ($\mu$m) & 0.00	& 0.00 & 0.00	& 0.00	& -6.48	& -1.29 & 2.90\\ \cmidrule(r){2-9}
 		& $\varepsilon$ ($\mu$m) & 20.06 & 22.12	& 25.11	& 22.47	& 18.84	& 21.52 & $\delta=4.10$\\ \midrule
\multirow{4}{*}{VT2\hspace{0.4in}} & $x$ ($\mu$m) &-17.21&-21.12&-24.57&-16.74&-11.33&-18.19&4.99\\
 		& $y$ ($\mu$m) & 17.85	& 6.45 & -1.14 &	-10.91 &	-9.21	& 0.61 & 11.87\\
 		& $z$ ($\mu$m) & -36.38	& 22.86	& 0.00	& 0.00	& 0.00	& -2.70 & 21.27\\ \cmidrule(r){2-9}
 		& $\varepsilon$ ($\mu$m) & 44.03 &	31.78	& 24.59	& 19.98	& 14.59	& 18.40 & $\delta=24.86$\\ \bottomrule
\end{tabular}
\end{center}
\end{table}
While the performance of the two virtual tetrodes was similar, as can be seen by the overlapping estimated source locations in Fig.~\ref{fig:vt}, VT1 was slightly less accurate, but substantially more precise, than VT2. However, in both cases, the estimates suffered from a relatively large bias.   

The presence of the bias indicates that the forward model defined by Eqs.~(\ref{eq:model}) and~(\ref{eq:lfv}) is inconsistent with the measurements. The cause of this discrepancy is elaborated on in Section~\ref{sec:d_icf}.
In the simplest scenario, this discrepancy can be mitigated  by relaxing the constant conductivity constraint, as outlined in Section~\ref{sec:icf}. According to this procedure, the ICF of the sensor closest to the source, tip sensor 4, was chosen as $k_4=1$. Note that this choice is arbitrary, as conductivities are expressed relative to one another. While the ICF is defined as a time-dependent quantity [Eq.~(\ref{eq:icf})], its values are remarkably constant except at instances when $E\{\psi_i(t)\}$ is near zero or passes through zero~\cite{Lee2011}. These outliers can be removed by taking the median value of $k_i$ over time. Following this procedure, the remaining ICFs were calculated as: $k_1=0.92,\, k_2=1.05,\, k_3=0.78,\, k_5=0.75$. These values can be interpreted as relative conductivities; the conductivity of tip sensors 2 was higher than that of tip sensor 4, while those of tip sensors 1, 3 and 5 were lower. The collected signals were then corrected by their corresponding ICFs (Sec.~\ref{sec:icf}), and MUSIC source localization was repeated using the corrected signals.  

The performance of the MUSIC algorithm improved significantly after scaling signals to correct for medium inhomogeneity (Table~\ref{tab:vt_after}). For VT1, the range of estimated source errors was reduced from $\sim$$[19,\,25]$ $\mu$m to $\sim$$[1,\,25]$ $\mu$m. In addition, a right-tailed t-test indicates that the mean error before signal correction was significantly higher than the mean error after signal correction ($\text{p-value} = 0.0163$). As in the before-correction case, there was no significant correlation between the error and the stimulation amplitude. After ICF correction, the average estimated source location (blue square in Fig.~\ref{fig:vt}) for VT1 was only 9.23 $\mu$m away from the source (down from 21.51 $\mu$m), with a standard radius of 10.70 $\mu$m (up from 4.10 $\mu$m). The improvement for VT2 was even more significant, as the error range was reduced from $\sim$$[15,\,44]$ $\mu$m to $\sim$$[1,\,4]$ $\mu$m. A right-tailed t-test confirmed that the reduction in error upon ICF signal correction was significant ($\text{p-value}= 0.00058$). In addition, no significant correlation was found between the error and the stimulation amplitude. The average estimated source location (yellow square in Fig.~\ref{fig:vt}) was only 1.30 $\mu$m away from the source (down from 18.40 $\mu$m), with a standard radius of 1.87 $\mu$m (down from 24.86 $\mu$m).   
\begin{table}[!htbp]%
\caption{Estimated source locations based on VT1 and VT2 data after correcting for inhomogeneity. 
The table setup is identical to that of Table~\ref{tab:vt_before}.}
\label{tab:vt_after}
\begin{center}
\begin{tabular}{crrrrrrrr}
\multicolumn{2}{c}{Stimulus amp. (V)} & 0.5 & 0.7 & 1.0 & 1.2 & 1.5 & Ave. & Std. dev.\\ \toprule
\multirow{4}{*}{VT1\hspace{0.4in}} & $x$ ($\mu$m) &0.15	& 3.24& -1.01	& 0.21 & -0.96	& 0.33 & 1.73\\
 		& $y$ ($\mu$m) & 2.14	& 10.94	& 0.25	& 1.08	& 3.78	& 3.64 & 4.29 \\
 		& $z$ ($\mu$m) & 0.00	& 22.57	& 0.00	& 6.31	& 13.50 & 8.48 & 9.65\\ \cmidrule(r){2-9}
 		& $\varepsilon$ ($\mu$m) & 2.15 & 25.29 & 1.04 & 6.41 & 14.05& 9.23 & $\delta=10.70$\\ \midrule
\multirow{4}{*}{VT2\hspace{0.4in}} & $x$ ($\mu$m) &0.53	& 3.28	& -0.78	& 0.82	& 0.47&0.86&1.48\\
 		& $y$ ($\mu$m) & 0.85	& 0.82 & 0.10 &	-1.03 &	-0.56	& 0.97 & 1.13\\
 		& $z$ ($\mu$m) & 0.00	& 0.00	& 0.00	& 0.00	& 0.00	& 0.00 & 0.00\\ \cmidrule(r){2-9}
 		& $\varepsilon$ ($\mu$m) & 1.91& 4.14 & 0.79 & 0.84 & 0.54 & 1.30 & $\delta=1.87$\\ \bottomrule
\end{tabular}
\end{center}
\end{table}

In summary, correcting for inhomogeneity significantly improved the performance of the MUSIC algorithm, with both average estimated source locations being less than 10 $\mu$m away from the true source location. The solutions were also very precise ($\delta < 11$ $\mu$m).  As a typical neuron soma is 10-50 $\mu$m in diameter~\cite{Humphrey}, these results are promising for neural source localization applications. Specifically the method may be capable of resolving the locations of neighboring neurons. 
  
\subsubsection{Virtual Pentode}\label{sec:vp}
\begin{figure}[!htbp]
	\centering
		\includegraphics[width=0.5\textwidth]{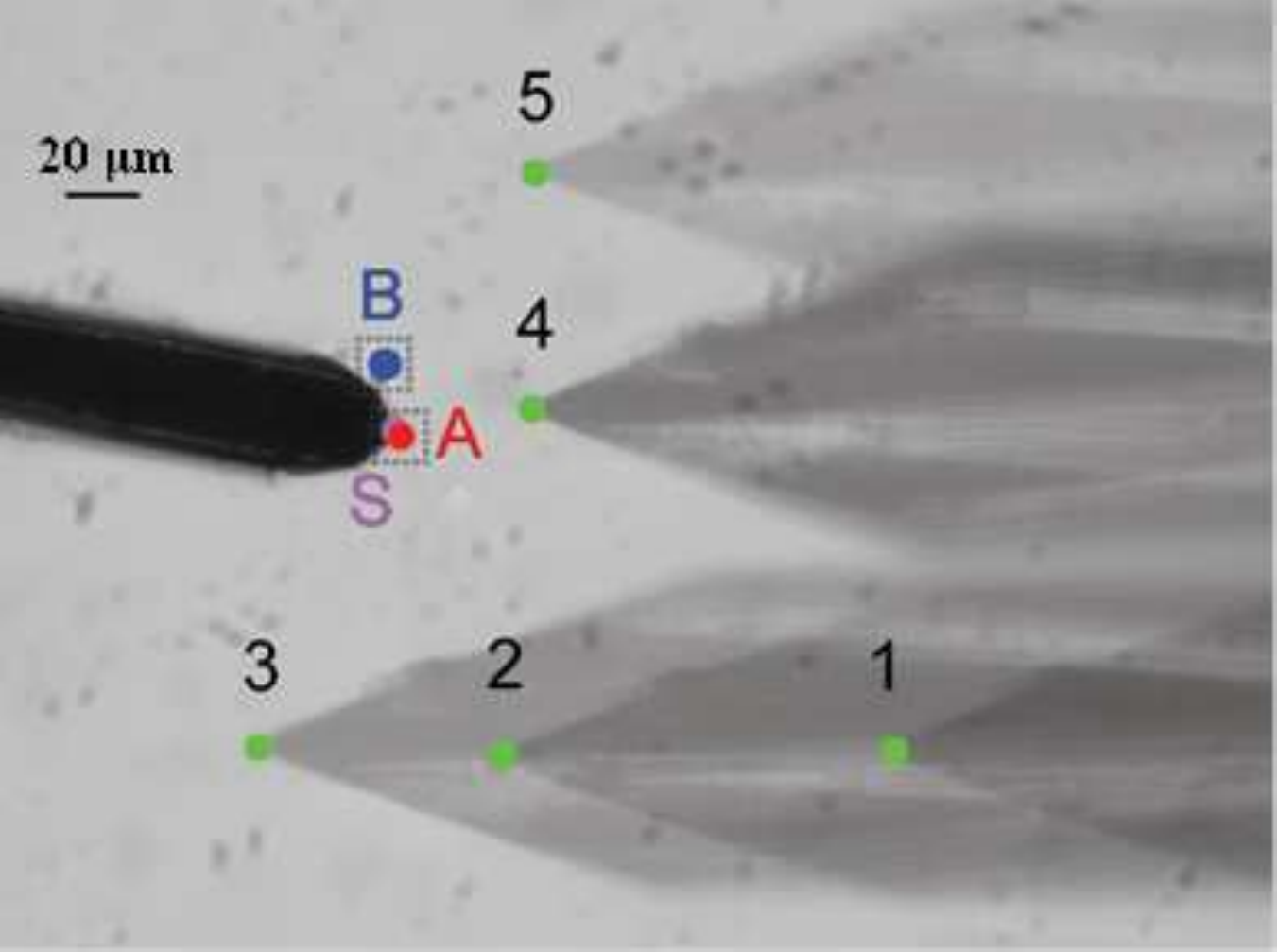}
	\caption{Superposition of five microscope images showing five tetrode positions (1 - 5) with tip positions marked by green circles. On the left is the stimulating electrode with the source, S, defined at the tip, and marked by a pink circle. The tetrode and stimulator were placed in the same focal plane for each depicted position. The blue and red circles represent VP MUSIC estimated source locations, averaged over 5 stimulus amplitudes, before (B) and after (A) ICF signal correction. The source location become more accurate after ICF correction. }\label{fig:vp}
\end{figure}

To investigate whether the addition of more sensors significantly impacts the localization results, a virtual array of all five sensors, the VP,  was also adopted for analysis (Fig.~\ref{fig:vp}). Table~\ref{tab:vp} depicts all localization results for the VP. Given raw signals, estimated source location errors ranged from $\sim$14 $\mu$m to $\sim$25 $\mu$m, and were very weakly, inversely correlated with the stimulation amplitude ($\rho = -0.58$, $\text{p-value} = 0.30$). The average estimated source location was 19.38 $\mu$m from the source (blue circle, B in Fig.~\ref{fig:vp}) with a standard radius of 7.37 $\mu$m (VP Before in Table~\ref{tab:vp}).

This result has a relatively large bias, which can be attributed to medium inhomogeneity. As in Sec.~\ref{sec:vt}, the signals were therefore corrected using derived ICF values. The resulting range of estimated source errors was reduced from $\sim$[14, 25] $\mu$m to only $\sim$[1, 4] $\mu$m. A right-tailed t-test indicated that the mean error after signal correction was significantly reduced from the mean error before correction ($\text{p-value}= 0.000004$). Likewise, after ICF correction the errors' inverse correlation with the stimulation amplitude became slightly more pronounced ($\rho = -0.68$, $\text{p-value}=0.21$), but still relatively weak. The average estimated source location, given corrected signals, was only 1.30 $\mu$m from the true source (red circle, A in Fig.~\ref{fig:vp}), with a standard radius of 1.88 $\mu$m (VP After in Table~\ref{tab:vp}), as compared with an error and standard radius of 19.38 $\mu$m and 7.37 $\mu$m, respectively, before correction.

\begin{table}[!htbp]%
\caption{Estimated source locations based on data acquired for the VP at different stimulation amplitudes.  The top row depicts results given raw data (VP Before), while the bottom row depicts results after signal correction for medium inhomogeneity (VP After).}
\label{tab:vp}
\begin{center}
\begin{tabular}{crrrrrrrr}
\multicolumn{2}{c}{Stimulus amp. (V)} & 0.5 & 0.7 & 1.0 & 1.2 & 1.5 & Ave. & Std. dev.\\ \toprule
\multirow{4}{*}{VP Before\hspace{0.4in}} & $x$ ($\mu$m) & -20.07 & -22.03 & -24.62 & -16.93 & -11.50 & -19.03 & 5.06\\
 		& $y$ ($\mu$m) & 0.82	& 0.73 & -0.98  & -10.21 &	-8.70	& -3.67 & 5.36 \\
 		& $z$ ($\mu$m) & 0.00	& 0.00 & 0.00	& 0.00	& 0.00	& 0.00 & 0.00\\ \cmidrule(r){2-9}
 		& $\varepsilon$ ($\mu$m) & 20.09 & 22.04	& 24.64	& 19.77	& 14.42	& 19.38 & $\delta=7.37$\\ \midrule
\multirow{4}{*}{VP After \hspace{0.4in}} & $x$ ($\mu$m) & 0.48 & 3.35 & -0.80 & 0.80 & 0.43 & 0.85 &1.52\\
 		& $y$ ($\mu$m) & 1.85	& 2.46 & 0.12 &	0.17 &	0.30	& 0.98 & 1.10\\
 		& $z$ ($\mu$m) & 0.00	& 0.00& 0.00	& 0.00	& 0.00	& 0.00 & 0.00\\ \cmidrule(r){2-9}
 		& $\varepsilon$ ($\mu$m) & 1.91 &	4.16	& 0.81	& 0.82	& 0.53	& 1.30 & $\delta=1.88$\\ \bottomrule
\end{tabular}
\end{center}
\end{table}

Overall, it appears that using an additional sensor does not necessarily imply an improvement in localization results. The VP performed better than VT1, but comparably to VT2. 

\subsection{Source Current Amplitude Estimation}
\begin{figure}[ht]
	\centering
	\begin{subfigure}[b]{0.455\textwidth}
		\centering
		\includegraphics[width=\textwidth]{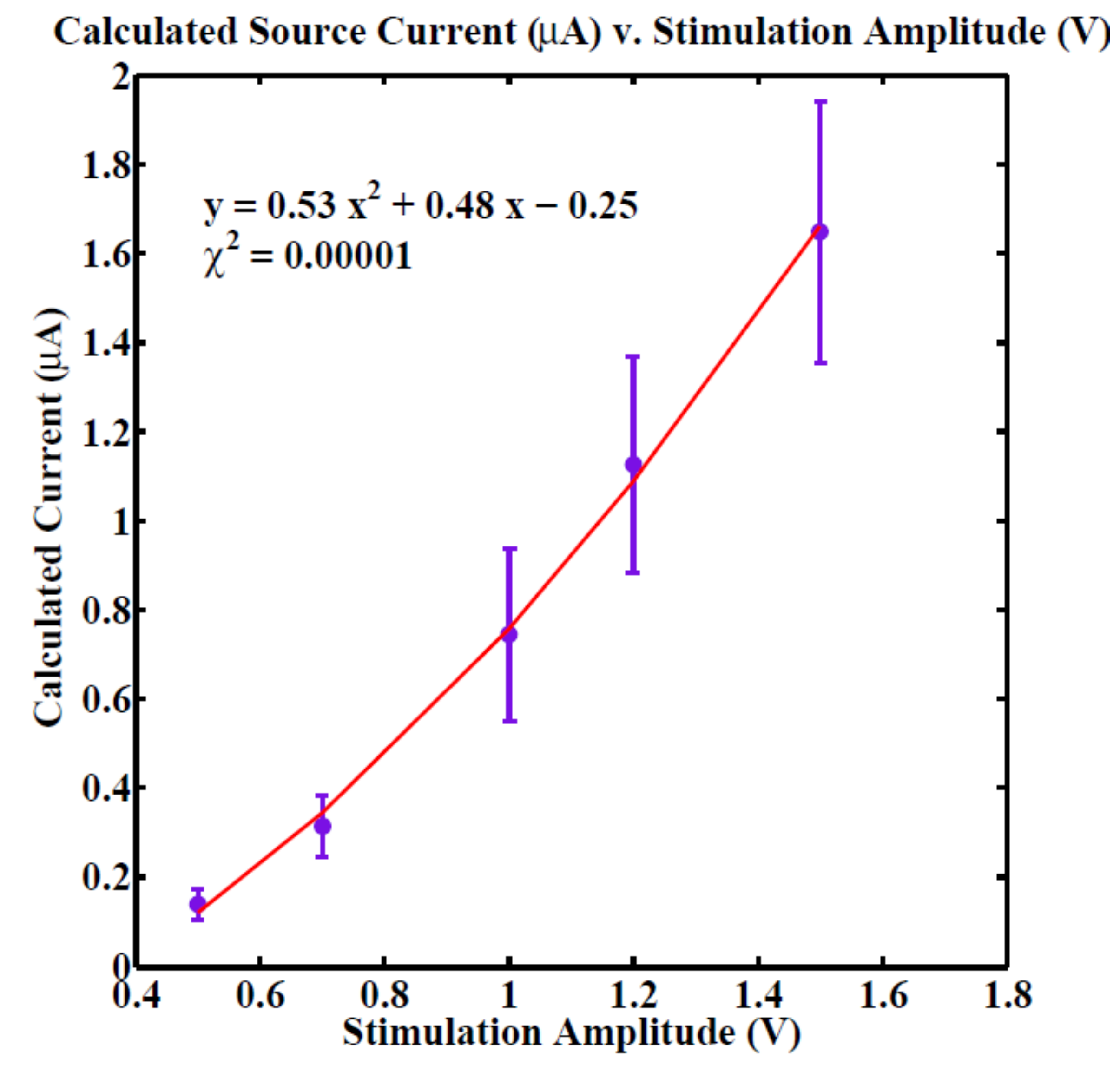}
		\caption{} 
		\label{fig:cva}
	\end{subfigure}
	\begin{subfigure}[b]{0.445\textwidth}
		\centering
		\includegraphics[width=\textwidth]{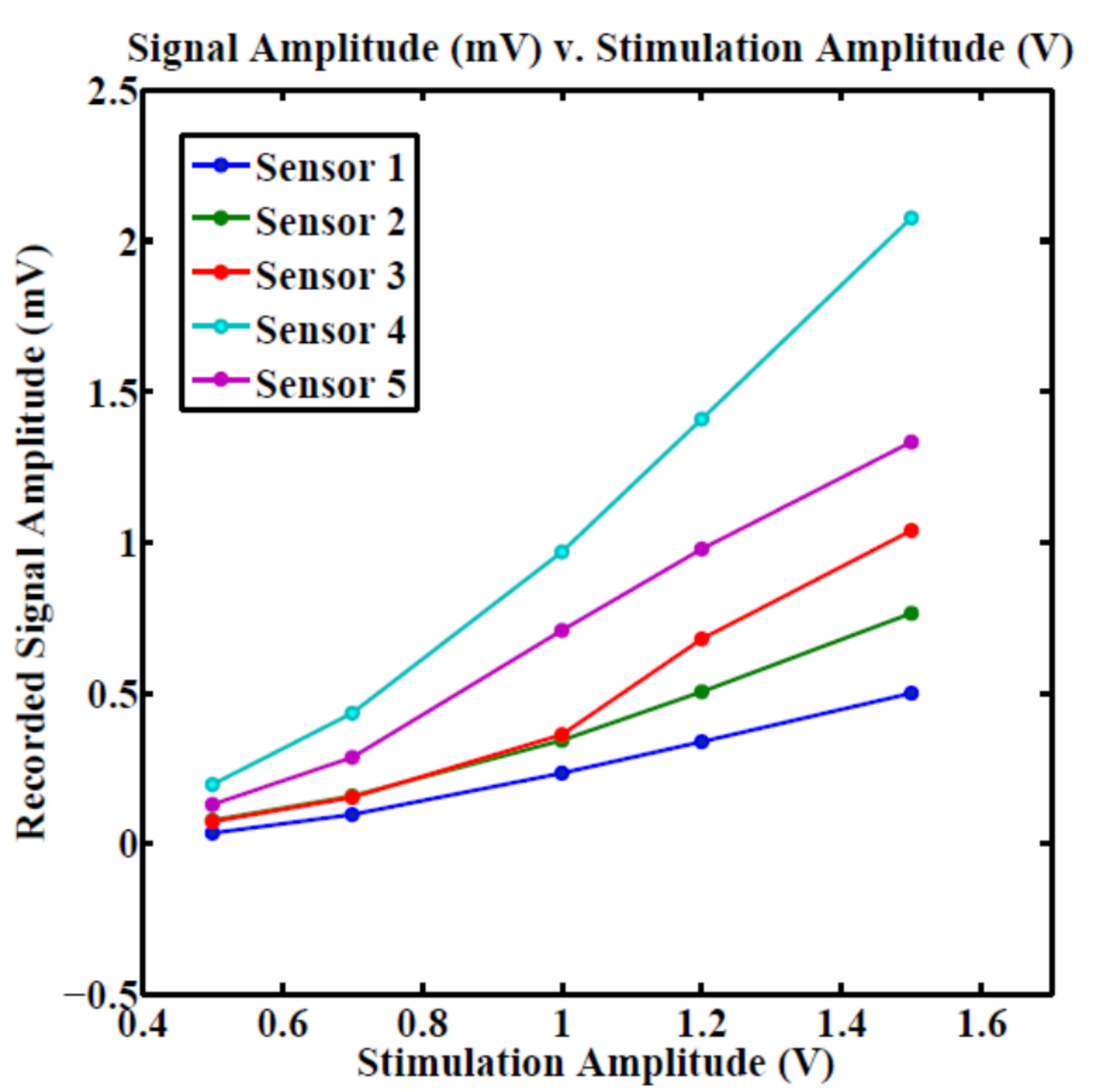}
		\caption{} 
		\label{fig:sva}
	\end{subfigure}

	\caption{(a) Estimated current amplitude ($\mu$A) as a function of stimulation amplitude (V). Each data point is an average of estimated currents across five tetrode tip positions. The data was fit to a polynomial: $y = 0.53x^2+0.48x-0.25$ ($\chi^2 = 0.00001$). (b) Recorded signal amplitude ($\mu$V) as a function of stimulation amplitude (V). Note that the signal amplitudes for each sensor also follow a quadratic trend, and has a negative y-intercept. This agrees with the estimated current results in (a), and implies that the phenomenon is not caused by our method but rather by a signal transformation occurring between the stimulator, medium, and sensors.}
\label{fig:current}
\end{figure}

Source current amplitude was estimated given ICF corrected signals and the corresponding VP localization results. Rearranging Eq.~(\ref{eq:expectation_val}) to solve for $I(t)$, given the five sensors, yields 
\begin{equation} \label{eq:current}
I(t) = E\{\psi_i(t)\} \times {4\pi k_i\sigma d_i}, \quad i = 1,\,2,\,\cdots,\,5
\end{equation}
where $d_i$ represents the distance between the estimated source location and sensor $i$, and $\sigma = 1.5 \times 10^{-6}$ S/$\mu$m, based on the ionic concentration of the medium. Although $I(t)$ can be calculated separately for each sensor $i$, these value should be consistent as there is only one true source. Likewise, although a time varying quantity, the current was only calculated given the peak value of $E\{\psi_i(t)\}$. The resulting $I^{*}$ then reflects the hypothesized peak current value, and was calculated for each stimulation amplitude.  The average $I^{*}$ over the five tetrode tip positions is shown as a function of stimulation amplitude in Fig.~\ref{fig:cva}. The estimated current was quadratically related to the stimulating amplitude, $y = 0.53x^2+0.48x-0.25$ ($\chi^2 = 0.00001$).
The same relationship, up to three significant figures, was also observed when $I^{*}$ was calculated using either VT1 or VT2 localization results. As shown in Fig.~\ref{fig:cva}, $I^{*}$ increased with stimulating amplitude, while the source location remained constant. Likewise, the precision of estimated currents, ranging from $ \pm 0.03$ $\mu$A to $ \pm 0.29$ $\mu$A (error bars in Fig.~\ref{fig:cva}) indicated that changes in source current can be detected. The algorithm's ability to estimate both source current and location, suggests it may be able to resolve changes in extracellular signal amplitude due to neuronal migration from changes in extracellular signal amplitude due to internal firing variations. 

Due to nonlinear interactions between the stimulating electrode and the medium, we cannot compare the estimated current at each stimulation amplitude to the true current being generated in the medium. However, a rough calculation using Ohm's Law, $V=IR$, indicates that given our stimulating amplitudes, and the corresponding estimated currents, the resistance of our setup is on the order of 1 M$\Omega$. This value agrees with the manufacturer specified impedance of our stimulating electrode, and is therefore a promising indicator that the estimated current amplitudes are accurate. 
It is expected that given no stimulation the resulting current should be zero, however the model in Fig.~\ref{fig:cva} indicates a negative y-intercept. When working with a dielectric solution such as saline, the response of the system to low stimulation is highly nonlinear~\cite{Zhang2011}, and the model can be expected to reflect this behavior given more data approaching zero. Additionally, note that a similar relationship was also observed when comparing the collected signal amplitude to the stimulation amplitude (Fig.~\ref{fig:sva}). This indicates that the observed behavior is not a consequence of our method, but rather a ramification of the signal transformation occurring at the stimulator, medium, and sensor interfaces. In general, our method accurately captures the behavior of the system, which is nonlinear especially at small stimulating amplitudes.  

\section{Discussion}\label{sec:d}

\subsection{MUSIC Source Identification}\label{sec:d_music}
According to the presented results, the MUSIC algorithm can accurately localize a source given a set of signals from four or more sensors. The estimated source locations proved to be accurate, less than 10 $\mu$m from the true source. As the diameter of a typical neuron soma is 10 - 50 $\mu$m~\cite{Humphrey}, these results imply that the MUSIC algorithm has the ability to successfully localize neurons. The precision of the solutions, with standard radii ranging from  1.87 $\mu$m to 10.70 $\mu$m, suggests the MUSIC algorithm may also be able to resolve the location of several closely spaced neurons. Furthermore, the algorithm's ability to accurately localize a source given five different stimulation amplitudes, indicates localization consistency even if the underlying signal changes. The proportional relationship between estimated source current amplitude and stimulation amplitude also showed that this method can accurately track changes in source amplitude. Note that the algorithm only relies on the relative signal amplitudes across sensors. Its performance will therefore not be affected given signals on the nA range typically observed {\it in vivo}~\cite{Jack}, as opposed to the $\mu$A range presented here. These results imply that the MUSIC algorithm can disambiguate a stationary neuron with varying firing amplitude from a migrating neuron with a steady firing pattern. Additionally, MUSIC may be able to resolve large distant neurons from small ones nearby. Our algorithm may therefore have significant implications for both acute and chronic electrophysiological studies. 

\subsection{Sensor Array Design Implications}\label{sec:d_sensor_design}
Virtual sensor array experiments indicated that increasing the number of sensors does not necessarily result in improved localization. The VP performed better than VT1, but comparably to VT2. As data from three sensors was shared between the VTs, the discrepancy between results from VT1 (average error = 9.23 $\mu$m) and VT2 (average error = 1.30 $\mu$m) can be attributed to their use of the sensors 1 and 3, respectively (Fig.~\ref{fig:vt}). The fact that VT2 performed significantly better than VT1, indicates that data from sensor 1 is inferior relative to the other four sensors. This could be due to experimental error, the sensor and stimulator's failure to be in the same focal plane, or the sensor's relatively large distance from the source. These results are not surprising given that the minimum number of sensors necessary to perform source localization is four, therefore we can expect the algorithm to be sensitive to outlying data in a tetrode arrangement. The VP, however, poses an overdetermined localization problem by using all five sensors. The resulting location estimate, with and average error of 1.30 $\mu$m, was relatively unaffected by the inclusion of sensor 1. This implies that although a superior localization result is not guaranteed with a higher number of sensors, arrays with a higher number of sensors are more robust against outlying data. With this in mind, microelectrodes with more than four sensors may be preferable for use in localization studies. 

\subsection{Calibrating Sensor Locations}\label{sec:d_sensor_locations}
Although for the purposes of validation, exact locations for the three circumferential sensors of a tetrode were not determined, the MUSIC algorithm can be used to establish these locations, as mentioned in Section~\ref{sec:es}. A setup similar to that presented in this paper, with a visible stimulator and tetrode tip, both of whose locations can be determined empirically, can be used to calibrate the remaining sensor locations. With the manufacturer's specifications the relative locations of the three circumferential sensors can be easily determined, the only unknown left being the tetrode's rotation about its own axis. MUSIC localization can then be performed for incrementally increasing tetrode rotation angles. This will result in a family of solution arranged in an oval, one segment of which will overlap with the true source location. The rotation angle responsible for the estimated source location closest to, or overlapping with, the true source location can then be accepted as the tetrode rotation angle. As the tetrode is not expected to change orientation while mounted in a holder or stereotactic device, this arrangement can be adopted as accurate throughout recording. We expect this kind of calibration to be performed prior to {\it in vivo} and {\it in vitro} recording, making the constraint of precisely knowing all sensor locations not an issue for neurophysiological experiments. 

\subsection{ICF Signal Correction}\label{sec:d_icf}
Our results show that adjusting for medium inhomogeneity significantly improved localization results. The inhomogeneity in our system is most likely coming from the presence of both the stimulating electrode and the recording tetrode whose dimensions are not negligible. This causes the spherical field symmetry characteristic of a monopole in a homogeneous isotropic medium to break. Furthermore the platinum-tungsten tetrode sensors generate concentration overpotential when immersed in an electrolyte, which changes the distribution of ions at the electrode-electrolyte interface. These sensors therefore behave as capacitors~\cite{Webster1997}, further violating the purely resistive monopole model presented in Eq.~(\ref{eq:monopolemodel}). As the assumption of a homogeneous medium is commonly made in neurophysiological studies~\cite{Holt1999}, the fact that this assumption does not hold, even in a simple saline medium, indicates that it may also be a factor during {\it in vivo} and {\it in vitro} experiments. Therefore, it may be worthwhile to explore adjusting existing techniques to take medium inhomogeneity into account. 

Although we believe medium inhomogeneity to be the main cause of the observed bias, ICF signal scaling could have also accounted for medium anisotropy, or varying sensor impedances. The virtual sensor array presented here used the tip sensor of a single electrode, making varying impedance an unlikely reason for the observed bias. However, medium anisotropy cannot be ruled out by this experiment.

Lastly, ICF adjustment was applied for purposes of validation only. As ICFs are derived using the known location of the source [Eq.~(\ref{eq:icf})], this adjustment cannot be performed on blind {\it in vitro} or {\it in vivo} collected data. We hypothesize that ICF signal correction can be made superfluous by prewhitening the signal prior to analysis. However, even if prewhitening does not mitigate an observed bias, this is not strictly an issue for neurophysiological experiments.  As long as individual neuronal sources can be disambiguated, their location bias relative to the recording electrode is irrelevant.  

\section{Conclusion}
The results presented here aptly demonstrate the MUSIC algorithm's ability to both localize and track the intensity of single sources. To the best of our knowledge this is the first experimental validation of electrical source localization and intensity characterization using multi-sensor arrays. Our localization accuracy, $<$ 10 $\mu$m from the source, is promising for future neuron localization studies {\it in vitro} and {\it in vivo}. Likewise, the precision of estimated source locations,  $<$ 11 $\mu$m, indicates that MUSIC can be used to resolve adjacent neuronal sources. Furthermore, estimated source intensity results imply that MUSIC can also be used to track changes in source signal amplitude. These outcomes present our method as a good candidate for source identification in both acute and chronic electrophysiological experiments, allowing much needed insight to be gained on neural migration patterns and size-related neuronal functionality.  

\section*{Acknowledgment}
This study was partially supported by the National Science Foundation
under Grant 1056105.

\bibliographystyle{ieeetr}
\bibliography{Amethodforneuralsourceidentification}

\end{document}